\documentclass[nonacm, sigconf,10pt]{_sty/acmart}
\pdfoutput=1
\usepackage{stfloats}
\usepackage[ruled,vlined]{algorithm2e}
\usepackage{url}
\usepackage{soul}
\usepackage{graphicx}
\soulregister\cite7
\AtBeginDocument{%
  \providecommand\BibTeX{{%
    \normalfont B\kern-0.5em{\scshape i\kern-0.25em b}\kern-0.8em\TeX}}}

\begin{document}

\title{FedHC: A Scalable Federated Learning Framework for Heterogeneous and Resource-Constrained Clients}


\author{Min Zhang}
\email{mzhang25@gmu.edu}
\affiliation{%
  \institution{George Mason University}
  \country{USA}}
  
\author{Fuxun Yu}
\email{fuxunyu@microsoft.com}
\affiliation{%
  \institution{Microsoft}
  \country{USA}}
   
\author{Yongbo Yu}
\email{yyu25@gmu.edu}
\affiliation{%
  \institution{George Mason University}
  \country{USA}}
   
\author{Minjia Zhang}
\email{minjiaz@microsoft.com}
\affiliation{%
  \institution{Microsoft}
  \country{USA}}
  
\author{Ang Li}
\email{angliece@umd.edu}
\affiliation{%
  \institution{University of Maryland}
  \country{USA}}

\author{Xiang Chen$^{\ast}$}
\email{xchen26@gmu.edu}
\affiliation{%
  \institution{George Mason University}
  \country{USA}}
\thanks{*Corresponding author}

\begin{abstract}

Federated Learning (FL) is a distributed learning paradigm that empowers edge devices to collaboratively learn a global model leveraging local data. Simulating FL on GPU is essential to expedite FL algorithm prototyping and evaluations. 
However, current FL frameworks overlook the disparity between algorithm simulation and real-world deployment, which arises from heterogeneous computing capabilities and imbalanced workloads, thus misleading evaluations of new algorithms. Additionally, they lack flexibility and scalability to accommodate resource-constrained clients.
In this paper, we present FedHC, a scalable federated learning framework for heterogeneous and resource-constrained clients. 
FedHC realizes system heterogeneity by allocating a dedicated and constrained GPU resource budget to each client, and also simulates workload heterogeneity in terms of framework-provided runtime. 
Furthermore, we enhance GPU resource utilization for scalable clients by introducing a dynamic client scheduler, process manager, and resource-sharing mechanism.
Our experiments demonstrate that FedHC has the capability to capture the influence of various factors on client execution time.
Moreover, despite resource constraints for each client, FedHC achieves state-of-the-art efficiency compared to existing frameworks without limits.
When subjecting existing frameworks to the same resource constraints, FedHC achieves a 2.75x speedup.
Code has been released on https://github.com/if-lab-repository/FedHC.

\end{abstract}




\maketitle
\section{INTRODUCTION}
\label{submission}



\begin{figure*}
    \centering
    \includegraphics[width=6.66in]{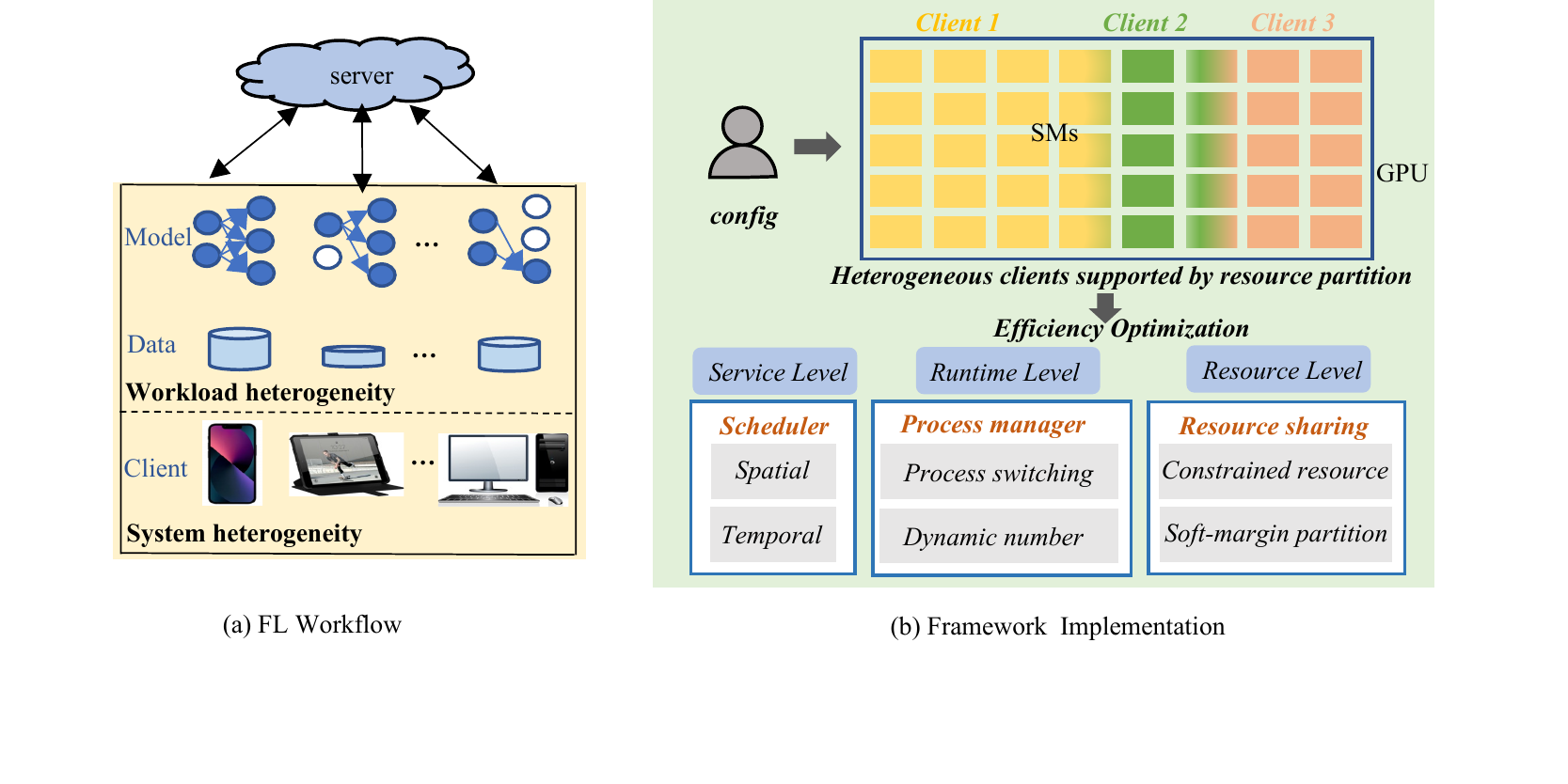}
    \vspace{-18mm}
    \caption{The high-level view of FedHC framework. (a) represents the FL workflow, which incorporates scalable clients with system and workload heterogeneity. (b) shows the implementation of FedHC, where varying proportions of computing units (SMs) on a single GPU are allocated to several parallel clients (three in this case), thus achieving system heterogeneity. Workload heterogeneity is reflected by framework-provided runtime. FedHC also enhances efficiency through optimizations at service level, runtime level and resource level.}
    \vspace{-3mm}
    \label{fig:FedHC_intro}
\end{figure*}

Federated learning (FL) emerges as a new distributed collaborative learning paradigm, which has drawn much attention from academia and industry \cite{mcmahan2017communication, hard2018federated, leroy2019federated, xu2021federated}.
Instead of centralizing data in a single server, federated learning allows data to remain decentralized on individual devices like smartphones. The model training process occurs on edge devices, with the model updates aggregated and sent to a central server for integration. 
In order to evaluate the performance of new FL algorithms before real-world deployment, FL simulation frameworks are designed using general CPU/GPUs to host a series of virtual clients, without the need of large-scale real edge devices.
FL simulation frameworks not only provide researchers with algorithm benchmarks but also efficient experimental environment, which helps researchers in FL community get started quickly and develop new FL algorithms.




Many existing FL frameworks \cite{caldas2018leaf, ingerman2019tensorflow, he2020fedml, beutel2022flower} focus on providing a platform to simulate different FL algorithms that tackle privacy, security, aggregation, and Non-IID data.
Besides, FedML \cite{he2020fedml} supports different computing paradigm, including hirarchical FL. Flower \cite{beutel2022flower} supports different programming languages and machine learning frameworks. Fedscale \cite{lai2021fedscale} supports communication heterogeneity and client availability simulation. 
However, current FL frameworks neglect the gap between algorithm simulation and real-world deployment, thus misleading evaluations of new algorithms. Additionally, they lack flexibility and scalability to effectively cater to resource-constrained clients. In the following, we describe the challenges of existing works and our proposed solution.





Firstly, the computation time of a client can be affected by many factors in realistic scenarios, such as hardware capability, data volume, model size, input sequence length, and batch size but existing FL frameworks often use oversimplified approach to estimate the client computation time. 
Slow clients caused by such factors train fewer local steps, resulting in low training quality. Aggregating models with varied training qualities slows down global model convergence.
Due to the lack of the accurate measurement mechanism for the computation time of the client on existing FL frameworks, when deploying algorithms on real-world edge devices, the heterogeneous computing capabilities and the diverse workloads will cause deviations of the global model convergence from the FL framework and the real world. 
Therefore, it is important to reflect the impact of heterogeneous factors on client's computation time.

There are three possible ways to represent the computation time of different clients but all of them are limited.
(1) Estimation by modeling the execution latency. This approach combines various factors that affect execution time into a single formula. FedScale \cite{lai2021fedscale} roughly estimated the execution time by the system speed and data volume, regardless of other factors such as model size, training configuration, input length, resource contention inside the client. It can not support the evaluation of some straggler acceleration algorithms working on reducing straggler's workload, such as resource-aware model personalization \cite{rapp2022distreal, deng2022tailorfl, bibikar2022federated, xu2021helios}. 
(2) Profiling models on real edge devices. Running the model directly on the real device yields realistic execution time. But this approach is time-consuming and labor-intensive due to variety of models and the need of large-scale edge devices. No FL framework currently uses profiling methods to achieve computational heterogeneity.
(3) Framework-provided runtime. This approach executes models on general CPU/GPUs and uses the wall-clock time on these backends as client's execution time. 
While it effectively handles workload heterogeneity, it lacks support for system heterogeneity within the same backend. For instance, clients sharing the same GPU possess identical computation capabilities, thus losing the ability to simulate diverse computational capacities.

To fully support the computation heterogeneity caused by system heterogeneity and workload heterogeneity, we propose a new method by assigning specific resource budgets to different clients for heterogeneous computing capabilities. Based on the resource budget, we further use framework-provided runtime to represent execution time.

Secondly, existing FL frameworks lack the resource management mechanism so that they fail to support scalable clients when considering resource constraints. They assume that clients have sufficient resources and do not impose any resource constraints on them. Based on this, 
LEAF \cite{caldas2018leaf} and TFF \cite{ingerman2019tensorflow} run different clients sequentially in single process, which is very time-consuming. Syft \cite{ryffel2018generic} and FederatedScope \cite{xie2022federatedscope} leverage distributed computing but require hardware nodes equal to the number of clients, which is extremely hardware-costly when scaling to massive clients. Flower \cite{beutel2022flower} and FedScale \cite{lai2021fedscale} support multiple clients on each hardware node but lacks resource management according to heterogeneous resource needs. 
To build an efficient FL framework featuring resource-constrained system heterogeneity, we face the following challenges:
(1) As a single GPU's supported parallelism is limited, the prevailing approach for simulating large-scale clients involves launching multiple parallel processes and sequentially running more clients within each process. Nevertheless, the CUDA context correlated with resource budget allocation, created within the process, cannot be altered, conflicting with our techniques for addressing system heterogeneity. 
(2) Existing frameworks set fixed number of processes which prevents the client's parallelism from adjusting based on resource usage.
(3) The total resource budget of parallel clients may not reach 100\%, as insufficient remaining resources for the subsequent client result in idleness. Therefore, a reasonable scheduling mechanism is needed to reduce resource waste between clients.
(4) Clients with substantial resource budgets may not maximize resource usage, leaving resources underutilized.


To overcome the limitation of heterogeneity and tackle the challenges of efficient framework, we propose \textit{FedHC}, a scalable federated learning framework for heterogeneous and resource-constrained clients. As shown in Fig.~\ref{fig:FedHC_intro}, our focus lies in enhancing computational flexibility and scalability. 
We summarize FedHC's five unique features as follows: 

\textbf{(\romannumeral1) Supporting system heterogeneity with constrained resource budgets.} To simulate the heterogeneous computational capabilities, we design a resource budget manager module that sets a maximum available percentage of GPU resource, representing various resource-constrained clients. 
We empower users to effortlessly configure the system heterogeneity by controlling resource partition.
\textbf{(\romannumeral2) Enabling workload heterogeneity in real runtime.} Besides hardware capability, we recognize that the running time of each client is also influenced by diverse workloads (caused by data volume, model size, input sequence length, and training configuration). As formulating an equation to accurately model or estimate these factors' impact on execution time proves challenging, we advocate utilizing platform-provided runtime as a representative of execution time. In doing so, diverse workloads are reflected in the runtime. 
\textbf{(\romannumeral3) Enhancing scalability via dynamic process management.} 
In order to solve the conflict between the re-used process and framework-provided system heterogeneity, and the problem of fixed parallelism, we propose dynamic process management.
We terminate the existing process after the client completes training and initiate a new process for allocating a new resource budget. Moreover, we devise a dynamic process management scheme, permitting additional processes during GPU resource idleness and fewer processes when GPU schedules are congested.
\textbf{(\romannumeral4) Optimizing resource utilization among clients via a scheduler.} 
To reduce the resource idleness among clients, we develop a client scheduler that orchestrates client execution order and parallelism based on resource budgets, utilizing a double-pointer selection module to identify the next client and a condition check module to determine deployment feasibility.
\textbf{(\romannumeral5) Improving resource utilization within client by resource-sharing parallelism.} 
To reduce the under-utilization of substantial resource budgets, we propose a resource-sharing method to tackle this problem.
Unlike conventional parallelism, our resource-sharing approach allows the cumulative resource budget of SMs percentage to surpass 100\% with clients competing for shared resources without breaching their individual maximum thresholds. This not only improves platform resource efficiency but also preserves system heterogeneity among clients.

We summarize the contributions of our work as follows:
\begin{itemize}
\item FedHC bridges the gap between simulation and real-world deployment with full consideration of system heterogeneity and workload heterogeneity. 
FedHC realizes system heterogeneity by allocating a dedicated and constrained GPU resource budget to each client, and also simulates workload heterogeneity in terms of framework-provided runtime. 

\item FedHC improves the GPU resource utilization through dynamic process management, client scheduling, and resource-sharing parallelism.
Flexible parallelism with optimization among and within resource budgets reduce resource idling.
It enables executing large-scale FL experiments under various heterogeneity configurations, even on a single GPU.

\item FedHC offers flexible APIs to extend its compatibility for both the hardware and software design. As the first FL framework to support explicit fine-grained resource management, FedHC enables the evaluation of heterogeneous model designs, resource optimization, and software-hardware co-design. We believe that FedHC will empower FL researchers and practitioners to explore a myriad of design opportunities concerning algorithms and resource optimizations.

\end{itemize}


\section{RELATED WORK}





Federated learning consists of various heterogeneous clients which collaboratively train a deep learning neural network.
Federated Learning faces four core challenges: Non-IID data distribution, privacy concerns, expensive communication, and computation heterogeneity.
Some works are proposed \cite{li2020federated, li2019convergence, yu2018parallel, huang2020loadaboost, jeong2018communication} to obtain convergence guarantees for Non-IID and unbalanced data. Methods like meta-learning and multi-task learning are extended to FL for modeling heterogeneous data \cite{smith2017federated, corinzia2019variational, eichner2019semi, khodak2019adaptive}.
Privacy-preserving approaches typically build upon classical cryptographic protocols like differential privacy \cite{bhowmick2018protection, mcmahan2017learning} and SMC \cite{bonawitz2017practical}. 
Model compression \cite{agarwal2018cpsgd}, split learning \cite{thapa2022splitfed}, and data compression techniques such as quantization and sketching \cite{alistarh2017qsgd, konevcny2016federated, ivkin2019communication} are proposed to reduce communication overheads. To tackle computation heterogeneity, asynchronous communication and active sampling of clients have been developed \cite{bonawitz2019towards, nishio2019client}. Model personalization also attracts much attention to reduce straggler's workload recently  \cite{rapp2022distreal, deng2022tailorfl, bibikar2022federated, xu2021helios}.


FL simulation frameworks are designed to expedite FL algorithm prototyping and evaluations before real-world deployment. They mainly focus on the FL features and platform efficiency.

From the aspect of FL features, most FL frameworks aims to establish benchmarks which integrate different datasets and algorithms \cite{caldas2018leaf, ingerman2019tensorflow, ryffel2018generic, he2020fedml}. But they neglects the heterogeneity of FL clients. 
Flower \cite{beutel2022flower} provided a fully language-agnostic interface through protocol-level integration, which supporting heterogeneous programming language and ML frameworks (Pytorch and Tensorflow). FederatedScope \cite{xie2022federatedscope} aims to support personalized FL config, flexible expression of behavior, and different loss function on local models. But both of they can not reflect the system heterogeneity and workload heterogeneity. 
FedScale \cite{lai2021fedscale} supported the system heterogeneity by providing a dataset of different computing speed. 
However, it can not support the workload heterogeneity caused by model size, input sequence length, data compression, and batch size. Overall, existing FL frameworks neglect the gap between algorithm simulation and real-world settings, thus misleading the evaluation of new algorithms.

From the aspect of framework efficiency, LEAF \cite{caldas2018leaf}, TFF \cite{ingerman2019tensorflow}, and FedML \cite{he2020fedml} run different clients sequentially on single hardware node, which is very time-consuming. Syft \cite{ryffel2018generic}, FederatedScope \cite{xie2022federatedscope} and FedML \cite{he2020fedml} support distributed computing but require hardware nodes equal to the number of clients, which is extremely hardware-costly when scaling to massive clients. Flower \cite{beutel2022flower} and FedScale \cite{lai2021fedscale} support multiple clients on each hardware node but lacks resource management according to heterogeneous resource needs. All of them can not support large-scale clients when applying framework-provided system heterogeneity.

\section{Heterogenous FL Implementation}
In this section, we briefly introduce the architecture of FedHC. Then we introduce how FedHC implements the system heterogeneity and workload heterogeneity.

\begin{figure}
    \centering
    \includegraphics[width=3.33in]{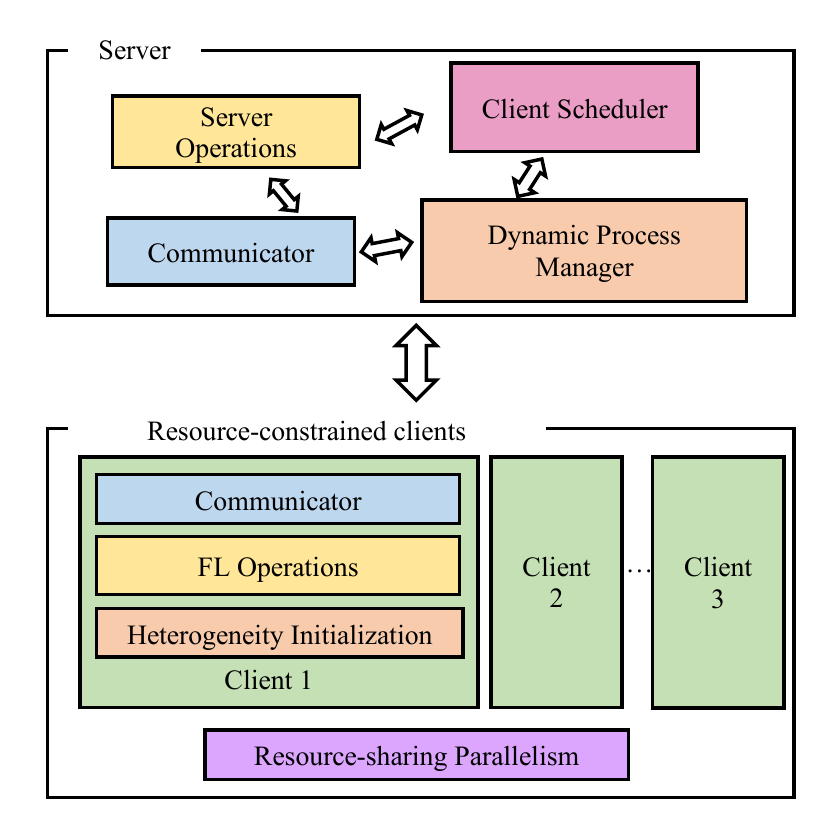}
    \vspace{-8mm}
    \caption{Architecture of FedHC framework}
    \label{fig:architecture overview}
    \vspace{-5mm}
\end{figure}

\subsection{Architecture Overview}
We design an FL framework, i.e., FedHC, to simulate the server and resource-constrained clients on GPU. Fig.~\ref{fig:architecture overview} illustrates the overview of FedHC architecture. The client scheduler determines the parallels clients to run, and then the dynamic process manager launches corresponding processes to execute these clients. The communicator facilitates the transmission of instructions and models between server and clients, while the resource-sharing parallelism module aims to boost GPU utilization. For each client, the corresponding process conducts the heterogeneity initialization to simulate the system heterogeneity. 

\subsection{FL Heterogeneity Simulation}

\noindent \textit{\textbf{Definition and impact of heterogeneous devices in FL:}}
From a computational perspective, client execution times result from heterogeneity in numerous aspects, which can be distilled into system and workload heterogeneity. 
System heterogeneity involves the different computing speed due to different hardware capabilities. 
Heterogeneous workload arises from data volume, model size, training configuration, and intermediate variables. 
Diverse training times lead to varying local model qualities, subsequently impacting the global model.
Therefore, simulating workload and system heterogeneity unveils disparities across clients in training times, further influencing aggregation and aiding researchers in developing and evaluating novel algorithms.


\noindent \textit{\textbf{Implementation of system heterogeneity:}}
Simulating system heterogeneity poses a challenge, as most simulation experiments lack numerous distinct hardware platforms. Running different clients on the same GPU results in homogeneous computation capability. To address this issue, we propose splitting GPU resources and allocating GPU resource shares to various clients. The differences in underlying resources available to each client yield varied computing speed, thus simulating the system heterogeneity.


\begin{figure}
    \centering
    \includegraphics[width=3.33in]{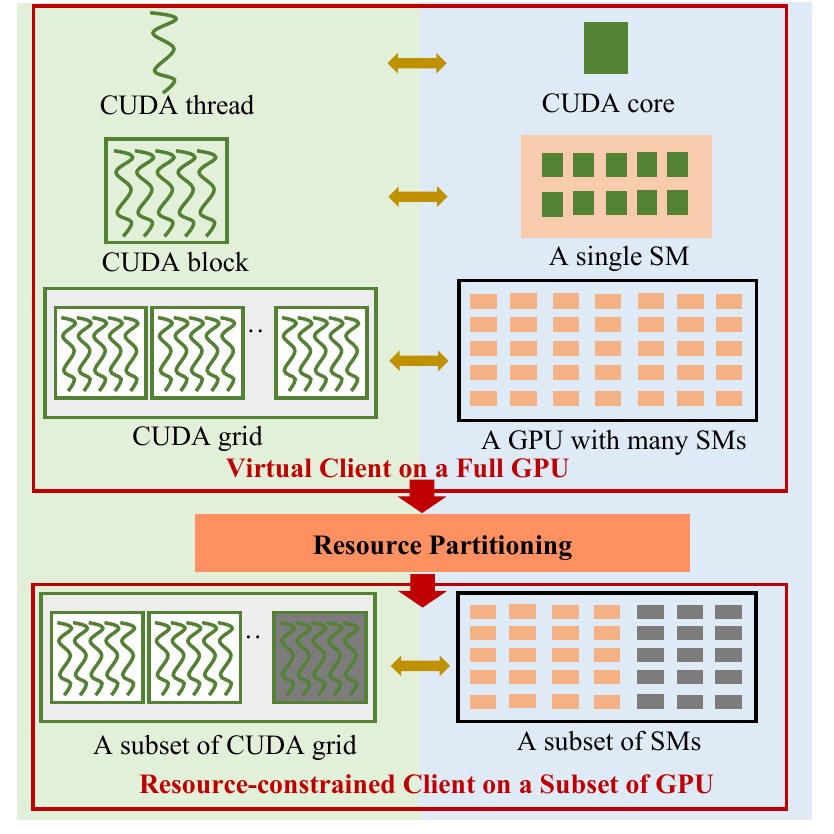}
    \vspace{-8mm}
    \caption{The implementation of heterogeneous computing capabilities. The correspondence of software (left) and hardware (right) for kernel execution on GPU shows how to implement the resource-constrained client by a subset of GPU.}
    \vspace{-5mm}
    \label{fig:fine-grain GPU}
\end{figure}

Fig.~\ref{fig:fine-grain GPU} illustrates how we implement the simulation of heterogeneous computing capabilities. 
When a model is deployed to execute on the GPU, it is composed of a series of CUDA kernels. The CUDA kernel is a function that get executed on GPU.
From a software aspect, a kernel is executed as a grid of thread blocks. When mapped to hardware execution, a CUDA block is executed by one streaming multiprocessor (SM) which consists of some CUDA cores. Depending on CUDA blocks' required resources, one SM can run multiple concurrent CUDA blocks. The CUDA grid comprised of several CUDA blocks is executed on many SMs. The number of SMs occupied depends on the size of the CUDA kernel. When the number of SMs is insufficient, the execution speed of the kernel will slow down. Based on this, our resource partitioning module simulates different computing speeds by limiting the number of SMs available. Instead of arbitrarily using any SMs on a full GPU, we set a \textit{\textbf{resource budget}} \textemdash  a subset of SMs, to simulate a resource-constrained client. Different computing capabilites can be obtained by setting different resource budgets. We implement the computing resource partitioning by \textit{os.environ["CUDA\_MPS\_ACTIVE\_THREAD\_PERCENTAGE"]}. The system heterogeneity we designed is very user-friendly, and only requires users to specify different resource budget parameters.


\noindent \textit{\textbf{Implementation of workload heterogeneity:}}
FedHC supports the simulation and evaluation of various workload heterogeneity. Users can flexibly configure imbalanced data volume or insert data compression method (i.e., data heterogeneity), design heterogeneous models via pruning and additional multi-task model (i.e., model heterogeneity), and customize hyper-parameters such as input sequence length and batch size in training (configuration heterogeneity). As these factors change, the corresponding alterations in training times and impact on global model are reflected. FedHC's ability to capture these changes stems from the deployments in real runtime. Instead of rough estimation, we record each client's wall-clock time as their training time. In synchronous aggregation, one global round's duration is the longest time among all clients' wall-clock times. In asynchronous aggregation, clients' participation in the current communication round is determined by their wall-clock times' order. 

Overall, FedHC users only need to set the percentage of computing units and workload-related configuration for each client. FedHC will constrain the resource allocated to the client according to the resource budget. By assigning different resource fractions, variations in running time can be incurred, thus realizing the simulation of system heterogeneity. At the same time, the real runtime of GPU has a natural advantage to reflect workload heterogeneity.

\section{Resource Optimization Implementation}

Heterogeneous clients have different needs of computing resource in realistic scenario. However, existing FL simulation frameworks ignore the feature so that they lack the resource management according to resource occupation. We find that existing FL framework can not support the efficient large-scale clients execution when applying our proposed heterogeneous FL simulation method. So we design the resource optimization to achieve the scalability in this section.


\subsection{Dynamic Process Manager}
The number of parallelism \textit{\textbf{m}} on single GPU is limited, but the number of participants \textit{\textbf{n (n>>m)}} of each global round in FL is massive. To achieve the scalability, we design a dynamic process manager, which contains process status monitor, process switching, process record table, and determination module.

\noindent \textit{\textbf{Challenge:}}
The current approach when running large-scale clients is to launch multiple parallel processes and run the client serially within each process. But it fails down when combined with system heterogeneity and also has the possibility of resource allocation waste or 'explosion'. As mentioned before, we implement the system heterogeneity by assigning a specific resource budget to a specific client at the start of the process. The resource limitation is maintained in the CUDA context. However, CUDA context within the process can only be created at the beginning of the process. Once the process has started, the CUDA context can not be changed, which means the resource budget within the process is constant. So the existing approach -- trying to sequentially run clients within each process -- can not support our proposed system heterogeneity. Moreover, existing approach can only keep fixed number of parallel processes. When the resource requirements of parallel clients are relatively small, the degree of parallelism should be increased. Conversely, parallelism should be reduced. 

\begin{figure}
    \centering
    \includegraphics[width=3.33in]{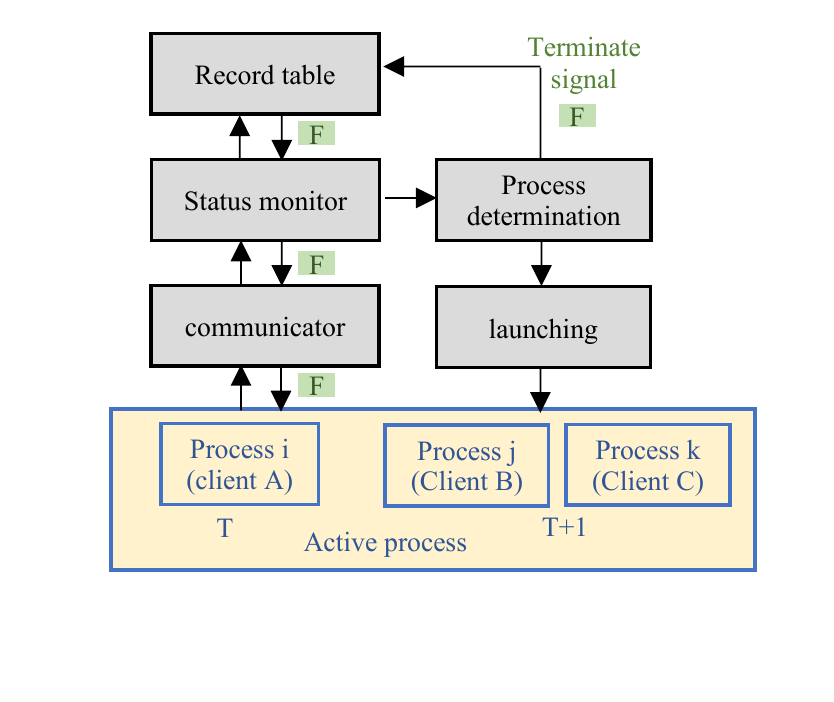}
    \vspace{-18mm}
    \caption{Dynamic process manager}
    \vspace{-5mm}
    \label{fig:dynamic_process_manager}
\end{figure}

\noindent \textit{\textbf{Method:}}
To tackle this challenge, we need to launch seperated process for each client for system heterogeneity consideration, and make the number of parallel processes adjustable according to resources needs. We propose a process manager as shown in Fig.~\ref{fig:dynamic_process_manager}. It has two main functions, process switching to achieve scalability and dynamic parallelism according to resource requirements.

The server is implemented in a long-lasting process, which is alive until the experiment ends. The server can dynamically launch processes for different clients. We use the Google Remote Procedure Call (gRPC) \cite{grpc2018high} to communicate between the server and clients. The communicator transmits the client's request to the server, and transmits the command signal generated by the server to the client. The status monitor processes the client's request and generates the next instruction. For example, when the state monitor receives the training completion signal from a client, it may issue an instruction signal for the client to upload its local model according to aggregation strategy. We also open APIs of testing handler, training handler, aggregation strategy for users to customize their settings. The status monitor stores the generated instructions in the record table. The record table has as many rows as the maximum number of parallel processes. Each row of the record table is a First-In-First-Out queue, containing the events to issue to the process. 

The processing switching to achieve scalability contains two aspects: terminating the old process and launching new process for the next client. For the aspect of terminating the old process, once the status monitor of the server receives the signal of client training completion, the process determination module will produce the terminate signal and save the signal in the record table of the corresponding row. It is transmitted to the client then. 
The client will jump out of the loop of continuously requesting the server when it receives the terminate signal. The process executing the client will be terminated. 
Another aspect of process switching is to launch new processes for subsequent clients. Once the next clients to be executed are determined, the launching module in the server will initiate a new process. The corresponding resource budget for system heterogeneity will be allocated at the beginning of the process, so that the resources available to the process cannot exceed the limit of the resource budget. In this way, we successfully solve the problem that clients running in the same process cannot satisfy system heterogeneity.

The process switching mechanism also breaks through the limitation of fixed parallel quantity, which allows dynamic change of parallel quantity.
Since the existing framework making the same process be reused by different clients, the number of clients running in parallel is always a fixed value. Unlike the fixed number of parallelism, our process switching is flexible to support dynamic number of parallelism. The process launching module can initiate any number of new processes under the premise that no process blockage occurs. The dynamic number of parallelism is determined by the scheduler which will be introduced in the next section. We use the limitation of total resource budget in scheduler and the parameter of maximum parallelism to avoid process blocking. The newly initiated process and the events to be issued for the new process will be recorded in the new queue in the record table.

\subsection{Client Scheduler}

\begin{algorithm}[t]
\caption{Resource-aware Scheduling Algorithm}
\label{alg:algorithm1}
\KwIn
{
$global$ resource budget list of running clients $running\_R$;

$global$ planned participants count $count$; 

$global$ available executor queue $AvailE$;

participant list $L$ containing client id and resource budget ($L_i$=($C_i, R_i$));

number of participants $N$;

total resource budget threshold $\theta$
}

\KwOut
{
to run client list $S$ containing client id, resource budget, executor id ($S_i$=($C_i, R_i, E_i$));
}

\BlankLine
Initialize $S$ $\gets$ $[]$

sorted $L$ according to resource budget $R_i$

\While{\textnormal{$count$ $<$ $N$ and $sum(running\_R)$ $<$ $\theta$}}
{
    current client $L_i$ $\gets$ left pointer 
    
    $stop\_flag$, $S$ $\gets$ \textbf{Check\_Current\_Client($L_i$, $S$)}

    \If{not ($count$ $<$ $N$ and $sum(running\_R)$ $<$ $\theta$)}
    {
    $stop\_flag$ $\gets$ True
    }
    
    \If{$stop\_flag$}
    {
    \Return{$S$}
    }
    
    current client $L_j$ $\gets$ right pointer 
    
    $stop\_flag$, $S$ $\gets$ \textbf{Check\_Current\_Client($L_j$, $S$)}
}
\Return{$S$}

\BlankLine
\BlankLine
\textbf{Check\_Current\_Client($L_i$, $S$):}

$stop\_flag$ $\gets$ False

\If{($R_i + sum(running\_R)$ $\leq$ $\theta$) and $AvailE$}
{
    
    
    {$E_i$ $\gets$ $AvailE.popleft()$}
    
    {$running\_R$.append($R_i$)}
    
    {$count$ $\gets$ $count+1$}
    
    $S$.append(($C_i$, $R_i$, $E_i$))
}
    
\Else
{
    \If{left pointer}{$stop\_flag$ $\gets$ True}
}
        
\Return{$stop\_flag$, $S$}
\end{algorithm}

As described in the previous section, when running large-scale clients, a combination of serial and parallel is required. 
Thus a scheduler related to temporal scheduling of client execution order and spatial scheduling of the amount of parallelism is necessary, so as to shorten the execution time of one global round and improve resource utilization.

\noindent \textit{\textbf{Challenge:}}
The current framework \cite{lai2021fedscale, beutel2022flower} adopts greedy scheduling. The selected participants in each global round are randomly arranged in the queue and they are scheduled by the order of the queue. The greedy scheduling can cause two problems.
Firstly, the low GPU utilization. When the current remaining GPU resources are less than the resource budget required by the next client, the next client cannot be deployed. As a result, the remaining GPU resources are wasted. So the scheduler should select appropriate parallel clients with the consideration of the resource budget.
Secondly, the long execution time of the global round. If the slow client is executed last and alone, the total execution time of a global round will be longer. Therefore, the slow client should try to have a higher priority when scheduling. However, if all the slow clients are executed at the beginning, the high parallelism may cause process blocking and has the risk of fragmentation problem. Moreover, the gathering of clients with large resource budgets leads to a decrease in parallelism, which cannot make good use of resources. So the scheduler should coordinate the order of clients for temporal consideration. 
Overall, the scheduler faces challenges of the spatial optimization to improve the GPU utilization and the temporal optimization to reduce the execution time of the global round.  

\noindent \textit{\textbf{Method:}}
To tackle these challenges, we propose a resource-aware scheduler with double pointers. When a client finishes executing, the server will call the scheduler to get the pending list of clients to run. As Algorithm.~\ref{alg:algorithm1} shows, we firstly sort the participants according to their resource budgets, and then cyclically use the left pointer and right pointer to alternately fetch the client until the end condition is met.
We use the left pointer to fetch the client with minimum resource budget, and use the right pointer to fetch the client with maximum resource budget. 
Once a client is selected to be executed, the condition checking module checks whether there are enough resources and idle processes to deploy the client. If the current client passes the conditional checks, the client is added to the pending list. On the contrary, when the remaining GPU resources are insufficient to sustain the client pointed by the right pointer, the right pointer will stop. But the left pointer will still continue, because the resource budget of the client on the left is less than that on the right, which can fill the remaining GPU resources. Until the client pointed by the left pointer cannot meet the condition checking, the algorithm will end and the pending list will be output. The pending list of clients with its resource budget and corresponding process will be used by the process launching module in the dynamic process manger as mentioned before.

In this way, clients with large and small resource budgets execute in parallel at the beginning, and clients with moderate resource budgets execute in parallel later. This prevents clients with small resource budget from slowing down the execution time of the global round, and also improves resource utilization.

\subsection{Resource-sharing Parallelism}

\begin{figure}
    \centering
    \includegraphics[width=3.33in]{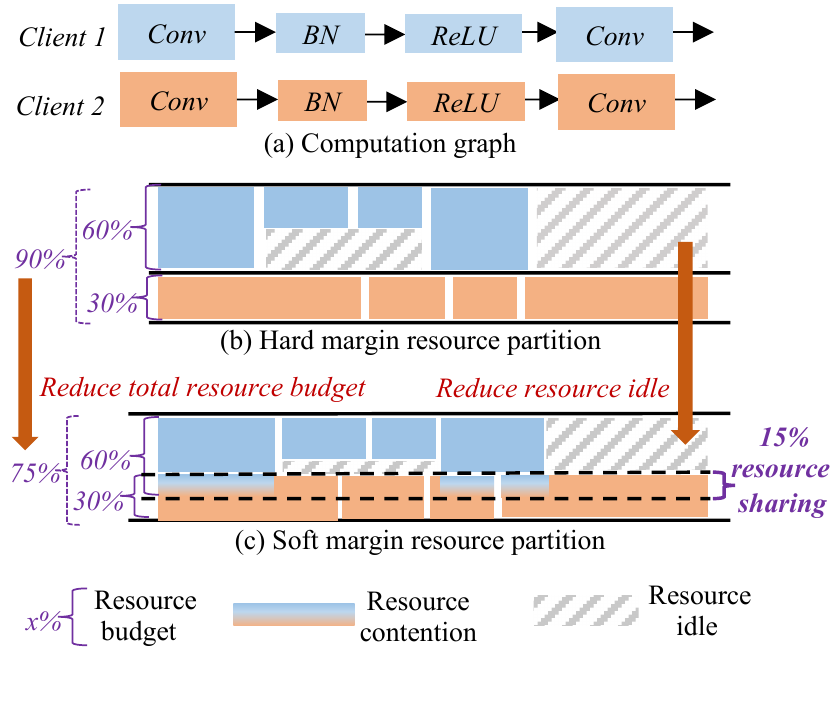}
    \vspace{-10mm}
    \caption{Design of resource sharing. We give a case of two parallel clients on a single GPU with two methods, hard margin resource partition (no resource sharing) and soft margin resource partition (resource sharing).}
    \vspace{-8mm}
    \label{fig:soft margin MPS}
\end{figure}

In order to host multiple clients running concurrently on single GPU, the server launches multiple processes at the same time. Each process is configured specific resource partition as mentioned in the previous section. We design two strategies of resource partition when clients are running in parallel. Fig.~\ref{fig:soft margin MPS} gives an example under these two strategies. Under the hard margin strategy, all clients execute within their own resource budget and do not affect each other. The soft margin strategy allows partial resource sharing. Although different clients compete to use some shared resources, but the amount of resources available to each client will not exceed its own resource budget limit.

Fig.~\ref{fig:soft margin MPS}(a) shows part of the computation graph of Client 1 and Client 2, containing operators such as convolution, batch normalization, and ReLU activation. When executing these two clients on single GPU, we assign 60\% and 30\% resource budget separately for each client so as to implement the hardware heterogeneity. Fig.~\ref{fig:soft margin MPS}(b) shows the execution of these two clients under the hard margin resource partition strategy. The resources occupied by the two clients are independent and will not affect each other. Different operators have different requirements for computing resources. Compute-intensive operators like convolution need more computing resource, otherwise it takes more time. When executing operators that don't require too many resources, there will be many idle resources in the large resource budget. These idle resources cause low GPU utilization.

To solve this problem, we propose soft margin resource partition strategy. As shown in Fig.~\ref{fig:soft margin MPS}(c), we allow 15\% computing resource sharing. For resource sharing area, two situation exist: Resource contention occurs when two big operators meet; If there are idle resources on one client, another client will fill the resource idle under the premise of not exceeding resource budget constraints.
Compared with hard margin resource partition strategy, the soft margin strategy has two advantages.
Firstly, the resource idle is reduced thanks to resource sharing, thus improving the GPU utilization.
Secondly, because of the resource overlap, the total resource budget is reduced from 90\% to 75\%, which leaves more resource space to increase the number of parallel clients. 

We use Multi-Process Service (MPS) \cite{MPS} to implement these two strategies.
FedHC provides the parameter to set the up-bounded resource constraint of all clients. Users only need to set the parameter at the beginning of the experiment. Hard margin resource partition strategy requires the parameter no more than 100\%. Otherwise, if the up-bounded resource constraint is set higher than 100\% (soft margin resource partition strategy), FedHC will automatically use the excess as a shared resource.

\section{FL Framework Comparison}




Table.~\ref{table:FL framework comparison}  summarizes the key differences between FedHC and existing FL frameworks.

\noindent \textit{\textbf{Heterogeneous Data}} means the data distribution among clients are Non-IID. The basic feature is supported by all frameworks.

\noindent \textit{\textbf{Heterogeneous Workload}} refers to the computation workload which is caused by several factors, such as data volume, data compression, model size, input sequence length and batch size. Existing frameworks(FedML, Flower, FedScale) only consider the unbalanced data volume, but neglects other factors. FedHC takes all of these factors into consideration.

\begin{table}[t]
\caption{Comparison with existing FL frameworks.}
\label{table:FL framework comparison}
\vspace{-3mm}
\begin{center}
\begin{normalsize}
\scalebox{0.8}{
\begin{tabular}{llccccc}
\toprule
Features                  & LEAF      & TFF       & FedML & Flower & FedScale & FedHC \\
\midrule
Heter. Data               & $\dagger$ & $\dagger$ & $\surd$   & $\surd$   & $\surd$   & $\surd$ \\
Heter. Workload          & $\times$  & $\times$  & $\dagger$ & $\dagger$ & $\dagger$ & $\surd$ \\
Heter. hardware      & $\times$  & $\times$  & $\times$ & $\dagger$ & $\dagger$ & $\surd$ \\
Resource optimization & $\times$  & $\times$  & $\times$  & $\dagger$  & $\dagger$  & $\surd$ \\
Scalability               & $\times$  & $\dagger$   & $\dagger$   & $\dagger$   & $\dagger$   & $\surd$ \\
Flexible APIs             & $\times$  & $\dagger$ & $\surd$ & $\surd$ & $\surd$ & $\surd$ \\

\bottomrule
\end{tabular}}
$\times$ means no support; $\surd$ means fully support; 

$\dagger$ means partially support.
\end{normalsize}
\end{center}
\vspace{-10mm}
\end{table}

\noindent \textit{\textbf{Heterogeneous Hardware}} is related to hardware computing capabilities. FedML and Flower support varying real edge devices, but acquiring a large-scale devices for an experiment is difficult. FedScale provides a dataset with different computing speeds, but limits it as a factor in the estimation formula. 
FedHC achieves the hardware heterogeneity by assigning different resource shares to different clients, which is more flexible and friendly.  



\noindent \textit{\textbf{Resource Optimization}} 
aims to improve the resource utilization and efficiency of the FL framework. Existing framework directly use the mechanism of the machine learning platform, ignoring the combination with resource-constrained clients in federated learning. 
FedHC conducts the resource optimization from the service level, runtime level and resource level.

\noindent \textit{\textbf{Scalability}} refers to execute large-scale clients in an efficient way.
TFF and LEAF are limited in single machine simulation. FedML support distributed computing but require hardware nodes equal to the number of clients. Flower and FedScale can simulate large-scale clients on a handful of GPUs, but they fail down when considering heterogeneous resource occupation. 
FedHC achieves the scalability of clients with different resource consumes.

\noindent \textit{\textbf{Flexible APIs}} allows the deployment and extension of diverse FL efforts. Apart from APIs such as data selection and model selection supported by existing frameworks, FedHC also provides heterogeneous resource initialization, explicit resource management, model variant, and personalized training configuration. 
\section{Experiments}


In this section, we evaluate FedHC's capabilities in the FL simulation experiments. Our evaluation focuses on two main aspects:

(1) Heterogeneous FL. We show that FedHC can simulate the system heterogeneity and workload heterogeneity with the framework-provided runtime. It can reflect the impact of different factors on client execution time that other frameworks cannot.

(2) Efficiency of the framework. We show that FedHC can effectively conduct large-scale FL experiments under various heterogeneity settings. We compare the efficiency with existing state-of-art FL frameworks.
Ablation experiments also show that the resource optimization components of FedHC (dynamic process manager, client scheduler, and resource-sharing parallelism) are effective to improve GPU utilization and efficiency. 

\noindent \textit{\textbf{Hardware environment:}}
All experiments are conducted using a single NVIDIA Titan V GPU. We assign different resource budgets to each client in order to simulate hardware heterogeneity. We execute each client with its respective resource budget and record the wall-clock time as the client's execution time.
\subsection{Support for Heterogeneous FL}
We firstly show that different factors could cause the training time of the client to be changed. Then, we show that how FedHC can use these factors to accelerate stragglers in FL but the state-of-art FL framework FedScale failed. Finally, we show the impact of the hardware heterogeneity and workload heterogeneity on the global convergence. 

\begin{figure}
    \centering
    \includegraphics{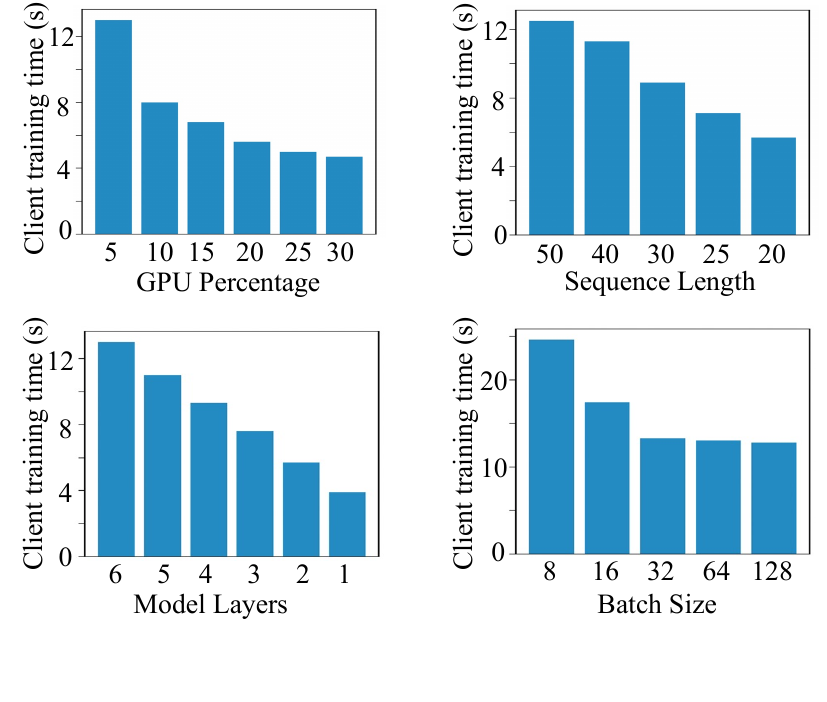}
    \vspace{-12mm}
    \caption{FedHC shows varied training time caused by diverse factors}
    \label{fig:impact}
    \vspace{-5mm}
\end{figure}

\begin{figure}
    \centering
    \includegraphics[width=3.3in]{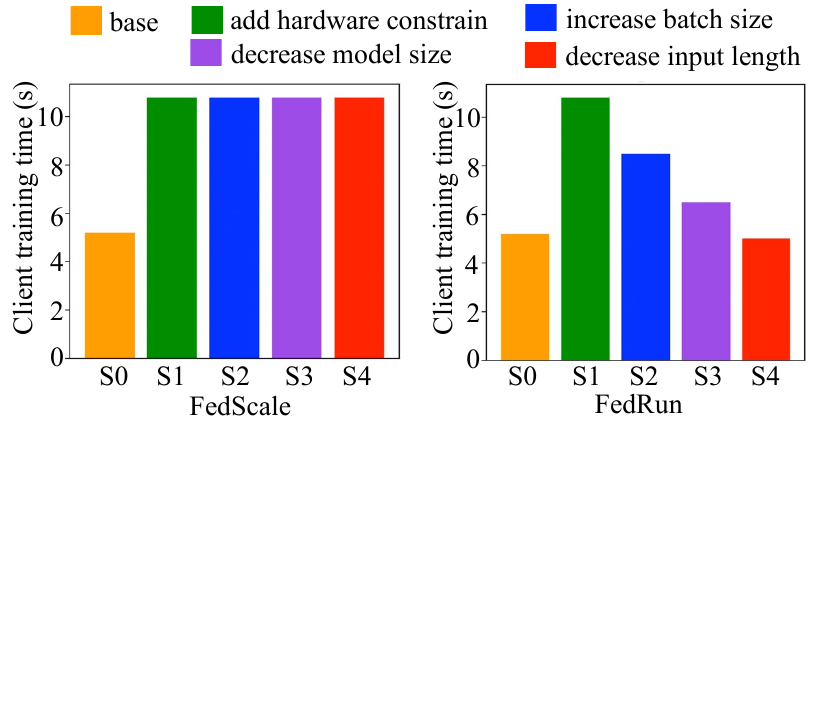}
    \vspace{-30mm}
    \caption{FedHC enables adjusting different factors to accelerate stragglers}
    \label{fig:straggler}
\end{figure}

\begin{figure}
    \centering
    \includegraphics[width=3.33in]{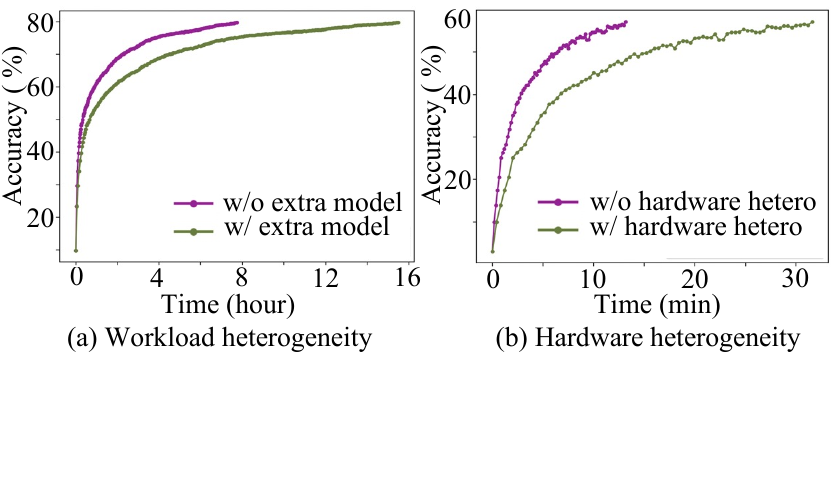}
    \vspace{-20mm}
    \caption{Impact of client heterogeneity on convergence}
    \label{fig:impact_convergence}
    \vspace{-8mm}
\end{figure}

\noindent \textit{\textbf{Experimental setup:}}
To show that FedHC can reflect the impact of different factors that could change the client training time, we change the value of resource budget, input sequence length, model layers, and batch size and record the client training time.
We experiment with the task of sentiment classification on the dataset SST-2 \cite{socher2013recursive}, the movie reviews with binary classes. We partition the data into Non-IID distribution. The base model we use is multi-layer LSTM. 

We also progressively adjust these factors on FedScale (the state-of-art FL framework) and FedHC to compare the heterogeneity support of different frameworks. 

Lastly, we design experiments to show the impact of the hardware heterogeneity and workload heterogeneity on the global convergence.
To show the effect of workload heterogeneity on the global convergence, we add an extra local model to increase the workload. Some works \cite{li2021ditto} are proposed to train a local model to perform client personalization, which increased workload for client. 
We train an image classification FL task on the Cifar10 \cite{krizhevsky2010cifar} dataset. 20 clients are generated in an Non-IID setting and 80\% clients are used to participate in local training in each global round. 
The accuracy of the global model over time is recorded separately with and without extra local model. 
To show the effect of hardware heterogeneity on the global convergence, we conduct experiments with/without hardware heterogeneity setting on FEMNIST dataset. In the setting without hardware heterogeneity, all clients are executed on the whole GPU. In the hardware heterogeneity setting, each client is assigned a specific resource budget to constrain the computing capabilities.

\noindent \textit{\textbf{Results:}}
Fig.~\ref{fig:impact} illustrates the client's training time varied under the impact of diverse factors. 
The smaller the GPU percentage, the longer the running time of the client, reflecting hardware heterogeneity.
The training time will be reduced when decreasing the input sequence length and number of model layers or increasing the batch size, which shows the workload heterogeneity.

Fig.~\ref{fig:straggler} shows that FedHC enables adjusting different factors to accelerate stragglers. The base model (S0) executed on GPU does not have any constraints of hardware capability and keeps the original workload. When adding hardware constraints setting (S1), the training time of the client on both FedScale and FedHC increases. However, when progressively changing other factors, i.e., increasing batching (S2), decreasing the number of model layers (S3), decreasing the input sequence length (S4), FedScale can reflect the reduced training time but FedScale fails. 

Fig.~\ref{fig:impact_convergence} shows the impact on convergence of workload heterogeneity and hardware heterogeneity. When adding the extra model, heavier workload occupies part of the client's resource, thus slowing down the speed of convergence. The convergence speed can also slowed down due to the hardware heterogeneity setting. Because clients with small resource budget have weaker computing capabilities, resulting in longer training time. 

With its ability to respond to different factors, FedHC avoids misleading the evaluation of client training time and convergence speed, thus narrowing the gap between simulation experiments and real-world deployments.


\begin{figure*}
    \centering
    \includegraphics[width=6.66in]{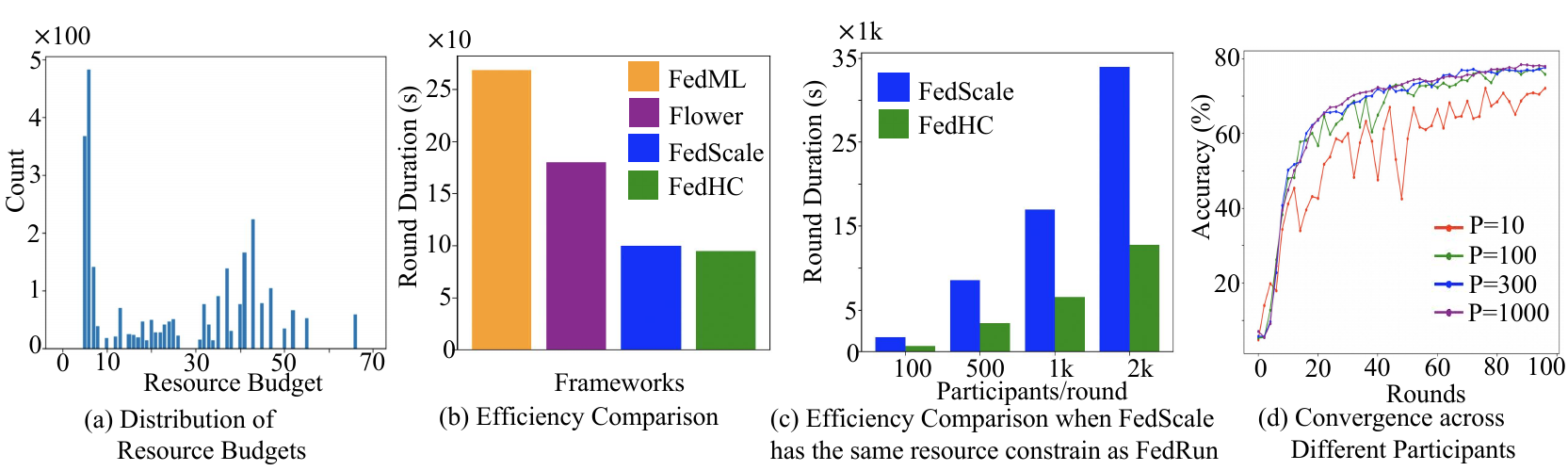}
    \vspace{-3mm}
    \caption{Result of scalability}
    \vspace{-3mm}
    \label{fig:result_scalability}
\end{figure*}

\begin{figure}
    \centering
    \includegraphics[width=3.33in]{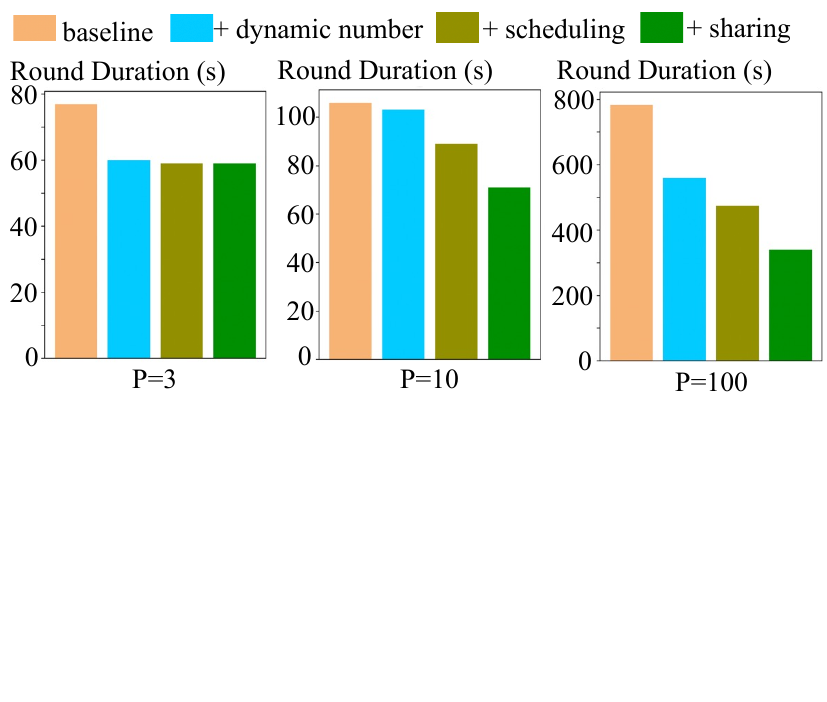}
    \vspace{-38mm}
    \caption{Ablation study}
    \vspace{-3mm}
    \label{fig:ablation_study}
\end{figure}

\subsection{Support for Scalability}
FedHC supports a large scale of clients to participant in the training process in each global round. Unlike existing FL frameworks, FedHC applies the constrained resource on each client. To show the efficiency of FedHC, we compare the round duration with existing FL frameworks.


\noindent \textit{\textbf{Experimental setup:}}
As FedHC assigns resource budgets to clients for the consideration of hardware heterogeneity. We transfer the computing speed dataset released in FedScale to the resource budget. We generate varying resource budgets for 2800 clients and illustrate the distribution in Fig.~\ref{fig:result_scalability} (a). The y-axis represents the number of clients with a specific resource budget, while the x-axis indicates the percentage of computing units on the GPU (SMs).
For framework comparison, we choose several advanced FL frameworks, FedML, Flower, and FedScale.
We use the FeMNIST dataset with a Non-IID partition and apply the same data to all frameworks for fair comparison.
We train the model ResNet18 with local data and aggregate models using FedAvg.
We set the same hyper parameters on all frameworks to keep the same workload. 10 clients are selected in one global round. For each client, we train on 500 batches of data, and the batch size is 64.
We record the duration including training, testing and so on for each round to compare the framework efficiency. 
We firstly use the original setting for each framework.
Next, we compare the framework efficiency in a more practical scenario where clients have limited resource in the context of FedScale, using the same setting as FedHC. We scale the number of participants in each round from 100 to 2000.
Finally, we evaluate the model convergence across different numbers of participants on FedHC framework.

\noindent \textit{\textbf{Results:}}
Fig.~\ref{fig:result_scalability} (b) illustrates the efficiency comparison with several advanced FL frameworks, including FedML, Flower, and FedScale. Despite FedHC having resource constraints for each client, unlike the other frameworks which have no limits, it still manages to achieve state-of-the-art efficiency when compared to existing frameworks. FedHC exhibits a slightly superior performance to FedScale.

However, when we apply the same resource-constrained client settings to FedScale, its efficiency significantly lags behind FedHC. As depicted in Fig.~\ref{fig:result_scalability} (c), FedHC achieves a 2.75x speedup compared to FedScale when the number of participant reaches 2000. This discrepancy arises from FedScale's lack of resource management for resource-constrained clients. Consequently, it cannot adaptively adjust client parallelism based on resource usage. Conversely, FedHC enhances GPU utilization by incorporating dynamic process management, resource-aware scheduling, and resource sharing mechanisms.

Fig.~\ref{fig:result_scalability} (d) shows the test accuracy across different number of participants. The global model achieves a faster convergence and reaches a higher accuracy when increasing the number of participants per round. Large-scale participants contribute larger data volumes and more diverse data distributions, resulting in higher model quality.

\begin{figure*}
    \centering
    \includegraphics[width=6.66in]{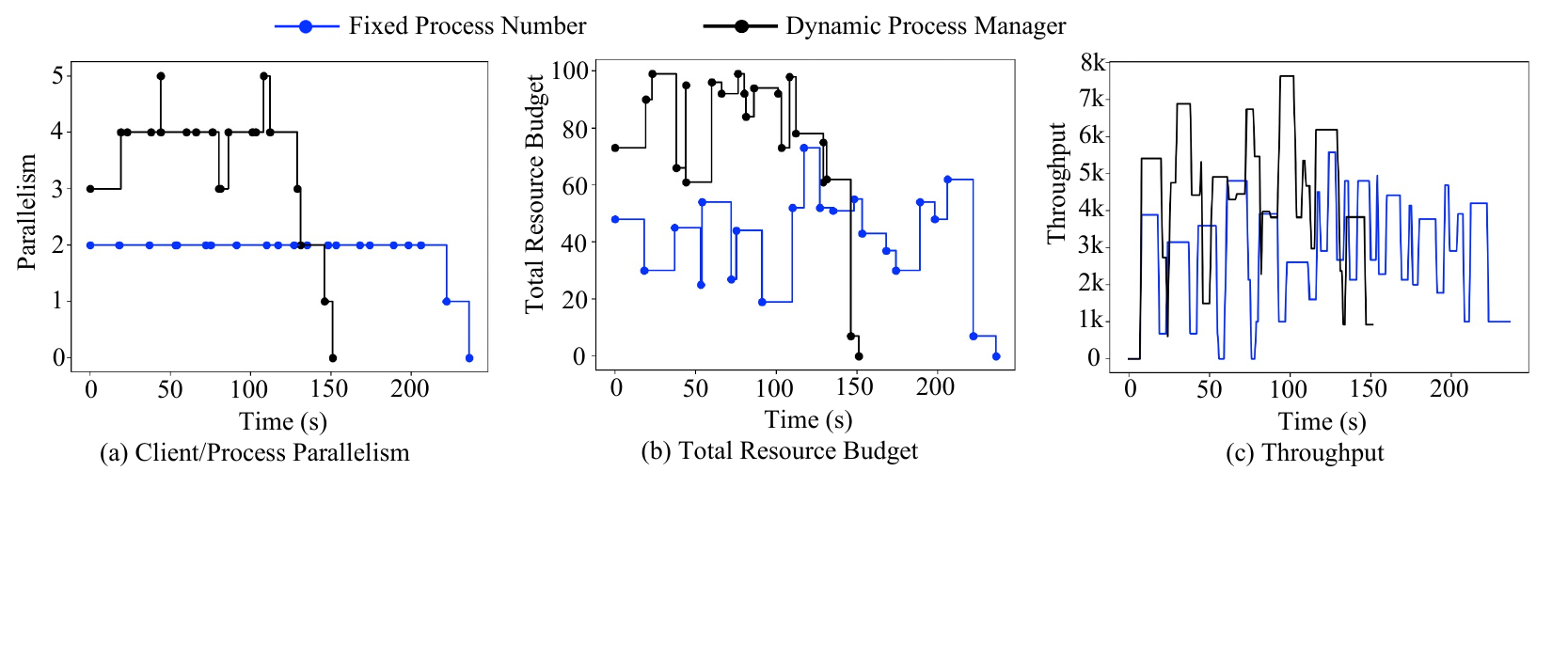}
    \vspace{-24mm}
    \caption{Performance under fixed process number and dynamic process number}
    \vspace{-3mm}
    \label{fig:parallel_dynamic_process}
\end{figure*}

\subsection{Effectiveness of FedHC components}

To demonstrate the effectiveness of each module, we design ablation experiments. Based on the FedScale structure framework, we added process switching to support the configuration of resource heterogeneity, which we used as the baseline for experimental comparison. We progressively add dynamic process management module, resource-aware scheduling module, and resource sharing module to realize FedHC. 
We use the same setting as last section. We select 3, 10, and 100 participants respectively, and report the execution time per global round. 

Fig.~\ref{fig:ablation_study} shows the execution time per global round with different number of participants, which can prove that each above modules in FedHC can reduce the execution time. Below we analyze the effectiveness of each module separately.

\subsection{Dynamic Process management}

Unlike the fixed number of parallel processes for client execution in FedScale, the dynamic process management in FedHC automatically determines the appropriate number of parallel processes based on the GPU resource usage.

Fig.\ref{fig:parallel_dynamic_process}(a) illustrates the variation in the number of parallel processes during a single global round involving a total of 20 participants. It is evident that the dynamic process management approach results in a higher and dynamic number of parallel clients compared to the fixed process number setting. As depicted in Fig.\ref{fig:parallel_dynamic_process}(b), this also translates into a clear advantage in terms of the total resource budget, ultimately leading to a reduction in execution time. This improvement is attributed to the resource management module's ability to adjust the level of parallelism based on the resource budget constraints of the parallel clients. In each global round, the resource management module proactively analyzes available resources and initiates additional processes when it predicts that there will be sufficient free resources. Fig.~\ref{fig:parallel_dynamic_process}(c) demonstrates that the throughput achieved with the dynamic process manager setting surpasses that of the fixed process number setting.

Moving on to Fig.~\ref{fig:kernel_dynamic_process}, we can visualize the parallelism of the kernel using the Nsight System tool. It is evident that under the fixed process number setting, the parallelism remains constant. However, in the dynamic process manager setting, the kernels execute with higher and varying degrees of parallelism.

\subsection{Scheduling}
The scheduler module in FedHC is responsible for determining the temporal execution order and spatial parallelism of participating clients, thereby further reducing the execution time of each global round.

\begin{figure*}
\centering
\begin{minipage}[t]{2.22in}
\centering
\setlength{\belowcaptionskip}{-15mm}
\includegraphics[width=2.22in,height=1.5in]{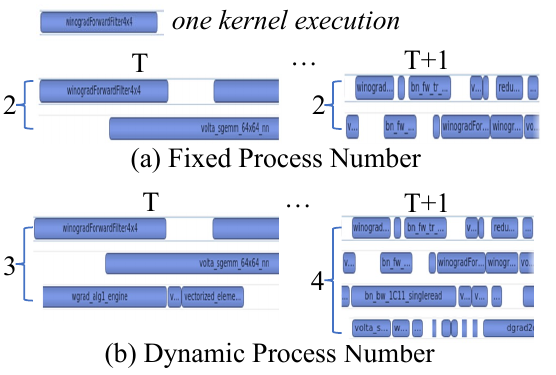}
\caption{Kernel parallelism}
\label{fig:kernel_dynamic_process}
\end{minipage}
\begin{minipage}[t]{4.44in}
\centering
\includegraphics[width=4.44in,height=1.5in]{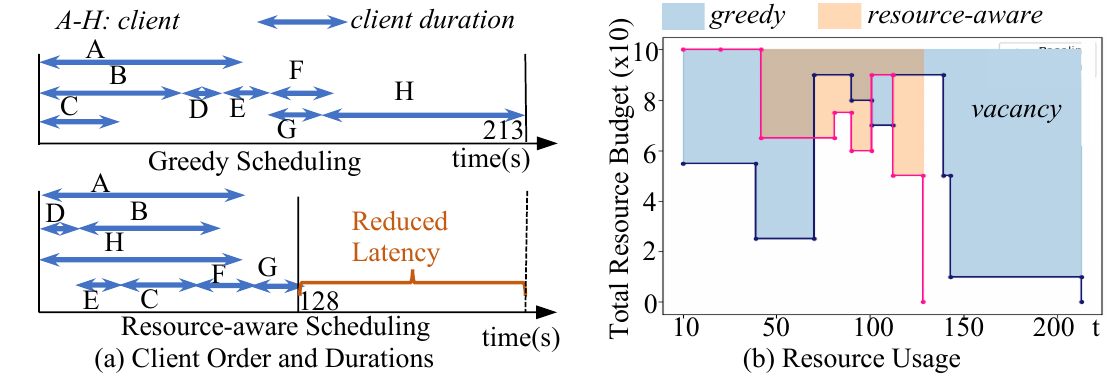}
\caption{Performance of resource-aware scheduling}
\label{fig:performance_resource_sche}
\end{minipage}
\end{figure*}

We present a case study involving 8 participants (A-H) randomly selected from a pool of 2800 clients. These participants have resource budgets of 10, 15, 30, 80, 65, 40, 50, and 10, respectively. In this study, we compare the outcomes of the existing greedy scheduling method with our resource-aware scheduling method.

In Figure~\ref{fig:performance_resource_sche} (a), the client order and execution durations are depicted. With the resource-aware scheduling, the order of execution for clients is altered, prioritizing the straggling client H to be executed earlier, thereby mitigating its impact on the overall duration. 
Furthermore, under the greedy scheduling setting, clients A, B, and C utilize 55\% of the computing resources, leaving insufficient resources on the GPU to accommodate client D, which requires an 80\% resource budget. 
Our approach enhances resource utilization and parallelism by coordinating the execution order of clients, effectively balancing resource-intensive and resource-light clients.
Consequently, the total execution time for a global round has been reduced from 213 seconds to 128 seconds.

Fig.~\ref{fig:performance_resource_sche} (b) shows the total resource budgets with different scheduling. The shaded region between the total resource budget curve and line y=100 represents vacant GPU resource that has not been assigned to any clients. Obviously, the resource-aware scheduling in FedHC has greatly reduced the the resource vacancy compared with the existing method.

\subsection{Resource Sharing}
Clients with substantial resource budgets may not maximize resource usage,
leaving resources underutilized.
FedHC uses the soft margin resource partition method, which allows clients to compete for the sharing resource part while each client does not exceed its own resource limit.
In this section, we show that resource sharing in FedHC improves the resource utilization, which achieves higher parallelism and throughput.

\begin{figure}
    \centering
    \includegraphics[width=3.33in]{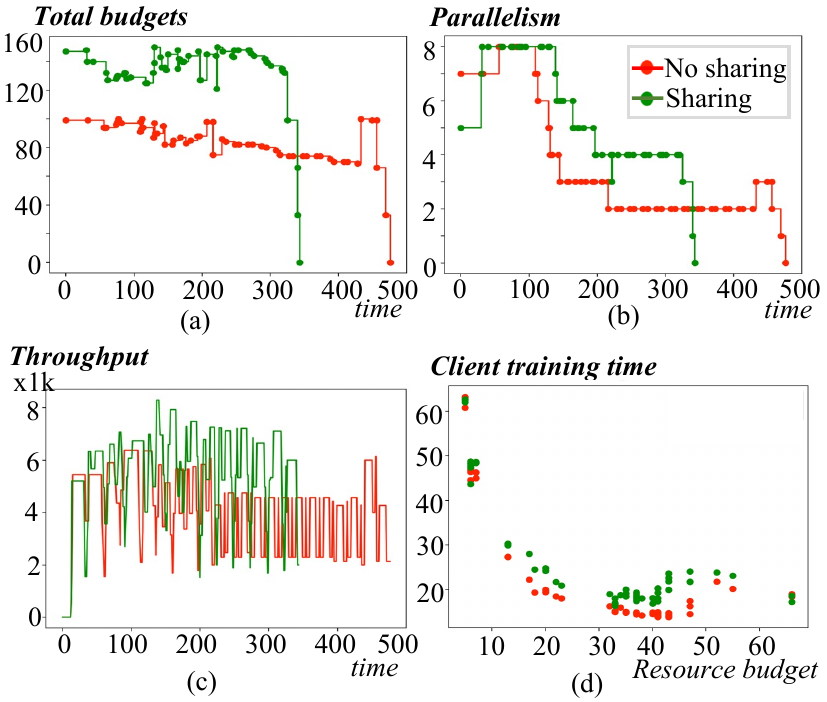}
    \vspace{-8mm}
    \caption{Performance of resource sharing}
    \label{fig:performance_of_sharing}
    \vspace{-5mm}
\end{figure}

\noindent \textit{\textbf{Experimental setup:}}
In hard margin resource partition setting, we set the total resource threshold as 100\%, which means no resource sharing. In soft margin resource partition setting, we set the total resource threshold as 150\%, which means 50\% GPU resource can be shared among parallel clients. We select 10 participants in each global round.

\noindent \textit{\textbf{Results:}}
Fig.~\ref{fig:performance_of_sharing}(a) shows the total resource budget of all participants in one global round. Resource sharing method improves the total resource budget so that the resource utilization can be improved. As a result, the total execution time of one global round is reduced. The resource sharing method improves the resource utilization by increasing the number of parallel clients as shown in Fig.~\ref{fig:performance_of_sharing}(b). That is because the resource sharing method makes full use of idle resources between resource budget. Fig.~\ref{fig:performance_of_sharing}(c) shows resource sharing method improves the throughput.
As Fig.~\ref{fig:performance_of_sharing}(d) shows, resource sharing also brings resource competition, resulting in variations in training time for each client. But we found the change to be small, especially for clients with small resource budgets. Therefore the total time of each global round is slightly affected because the straggler's time dominates.

\section{Conclusion}
To enable simulation of large-scale heterogeneous devices in real FL, we introduce FedHC, a scalable federated learning framework for heterogeneous and resource-constrained clients. Existing FL research platforms do not support scalable FL clients especially when considering hardware heterogeneity and diverse workloads. Unlike rough estimation methods, we assign a resource budget to each client, resulting in varying runtimes due to heterogeneous resource constraints, and any workload heterogeneity can also be reflected within the runtime. Furthermore, we enhance resource utilization of GPU for scalable clients through a dynamic process manager to control parallelism, a client scheduler for temporal and spatial coordination, and a resource-sharing method to reduce idling. Experiments show FedHC can perform large scale FL experiments with heterogeneous FL scenarios, enabling researchers to explore more FL design opportunities.

\newpage
\bibliography{_ref/main_paper}
\bibliographystyle{_sty/ACM-Reference-Format}

\end{document}